\documentclass[twocolumn]{aastex63}
\usepackage{graphicx}

\usepackage{pifont}
\renewcommand{\checkmark}{\ding{51}}%
\newcommand{\xmark}{\ding{55}}%

\begin{document}
\title{Nine localised deviations from Keplerian rotation in the DSHARP circumstellar disks:\\ Kinematic evidence for protoplanets carving the gaps}

\author[0000-0001-5907-5179]{C. Pinte}
\email{christophe.pinte@monash.edu}
\affiliation{School of Physics and Astronomy, Monash University, Clayton Vic 3800, Australia}
\affiliation{Univ. Grenoble Alpes, CNRS, IPAG, F-38000 Grenoble, France}
\author[0000-0002-4716-4235]{D.J. Price}
\affiliation{School of Physics and Astronomy, Monash University, Clayton Vic
3800, Australia}
\author[0000-0002-1637-7393]{F. M\'enard}
\affiliation{Univ. Grenoble Alpes, CNRS, IPAG, F-38000 Grenoble, France}
\author[0000-0002-5092-6464]{G. Duch\^ene}
\affiliation{Astronomy Department, University of California, Berkeley, CA 94720-3411, USA}
\affiliation{Univ. Grenoble Alpes, CNRS, IPAG, F-38000 Grenoble, France}
\author[0000-0002-0101-8814]{V. Christiaens}
\affiliation{School of Physics and Astronomy, Monash University, Clayton Vic 3800, Australia}
\author[0000-0003-2253-2270]{S.M. Andrews}
\affiliation{Harvard-Smithsonian Center for Astrophysics, 60 Garden Street, Cambridge, MA 02138, USA}
\author[0000-0001-6947-6072]{J. Huang}
\affiliation{Harvard-Smithsonian Center for Astrophysics, 60 Garden Street, Cambridge, MA 02138, USA}
\author[0000-0003-3253-1255]{T. Hill}
\affiliation{Atacama Large Millimeter/Submillimeter Array, Joint ALMA Observatory, Alonso de C\'ordova 3107, Vitacura 763-0355, Santiago, Chile}
\author[0000-0001-5688-187X]{G. van der Plas}
\affiliation{Univ. Grenoble Alpes, CNRS, IPAG, F-38000 Grenoble, France}
\author[0000-0002-1199-9564]{L.M. Perez}
\affiliation{Departamento de Astronom\'ia, Universidad de Chile, Camino El Observatorio 1515, Las Condes, Santiago, Chile}
\author[0000-0001-8061-2207]{A. Isella}
\affiliation{Department of Physics and Astronomy, Rice University, 6100 Main Street, MS-108, Houston, TX 77005, USA}
\author[0000-0002-8692-8744]{Y. Boehler}
\affiliation{Univ. Grenoble Alpes, CNRS, IPAG, F-38000 Grenoble, France}
\author{W.R.F. Dent}
\affiliation{Atacama Large Millimeter/Submillimeter Array, Joint ALMA Observatory, Alonso de C\'ordova 3107, Vitacura 763-0355, Santiago, Chile}
\author[0000-0002-5526-8798]{D. Mentiplay}
\affiliation{School of Physics and Astronomy, Monash University, Clayton Vic 3800, Australia}
\author[0000-0002-8932-1219]{R.A. Loomis}
\affiliation{NRAO, 520 Edgemont Rd, Charlottesville, VA 22903, USA}

\begin{abstract}
  We present evidence for localised deviations from Keplerian rotation, i.e., velocity ``kinks'', in 8 of 18 circumstellar disks observed by the DSHARP program: DoAr~25, Elias~2-27, GW~Lup, HD~143006, HD~163296, IM~Lup, Sz~129 and WaOph~6. Most of the kinks are detected over a small range in both radial extent and velocity, suggesting a planetary origin, but for some of them foreground contamination prevents us from measuring their spatial and velocity extent. Because of the DSHARP limited spectral resolution and signal-to-noise in the $^{12}$CO J=2-1 line, as well as cloud contamination, the kinks are usually detected in only one spectral channel, and will require confirmation. The strongest circumstantial evidence for protoplanets in the absence of higher spectral resolution data and additional tracers is that, upon deprojection, we find that \emph{all} of the candidate planets lie within a gap and/or at the end of a spiral detected in dust continuum emission. This suggests that a significant fraction of the dust gaps and spirals observed by ALMA in disks are caused by embedded protoplanets.
\end{abstract}


\section{Introduction}

Since the first spectacular ALMA images of HL~Tau \citep{ALMA_HLTau}, we have speculated about the origin of rings and gaps in disks. Proposed explanations include snow lines \citep[e.g.][]{Zhang15}, dust grain sintering \citep{Okuzumi2016}, non-ideal MHD effects and zonal flows \citep{Flock2015,Riols2019}, self-induced dust-traps \citep{Gonzalez2017}. The most tantalising explanation is that dust gaps are caused by embedded planets \citep{Dipierro15b,Jin2016,Dong2017,Zhang_DSHARP}

Direct imaging of putative planets in young circumstellar disks has proved difficult.
After several years of surveys using the new generation of adaptive optics instruments, the only confirmed directly imaged protoplanets are located in the gap/cavity of the transition disk around PDS~70 \citep{Keppler2018,Muller2018,Christaens2019,Haffert2019}.

A complementary approach is to search for kinematic signatures of planets.
Embedded planets perturb the Keplerian gas flow in their vicinity, launching spiral waves at Lindblad resonances both inside and outside their orbits \citep{Goldreich1979}. The disturbed velocity pattern is detectable with high spectral and high spatial resolution ALMA line observations.
Accurate measurements of rotation curves revealed for instance radial pressure gradients, likely driven by gaps carved in the gas surface density by Jupiter-mass planets in the disk of HD~163296 \citep{Teague2018}.
In a given channel map, the emission is concentrated along the iso-velocity curve, \emph{i.e.}, the region of the disk where the projected velocity is constant. In the presence of a planet, the iso-velocity is distorted and the emission displays a distinctive ``kink''. This technique led to the detection of embedded planets in the disks surrounding HD~163296 \citep{Pinte2018b} and HD~97048 \citep{Pinte2019}, with masses 2--3 times that of Jupiter. Similarly, deviations from Keplerian rotation  were detected in the disk of HD~100546 \citep{Perez2019}.

In all three cases, the velocity kinks coincide with a gap, demonstrating that protoplanets are responsible for at least some of the gaps observed in disks.
The recent spectacular series of ALMA high angular resolution campaigns
(\citealp{Long2018a,Huang2018a}, see also a collection of other datasets in \citealp{Marel2019}) have shown that rings and gaps are common, and they are now known in more than 30 disks. Are all of these gaps associated with protoplanets?

To answer this question, we have searched through existing ALMA archival data for kinks. Few datasets have the required signal-to-noise ratio, spectral and spatial resolution to reveal kinks, except the data from the DSHARP program \citep{Andrews2018b, Huang2018a}. We find nine candidate kinks.

\section{Observations and imaging}

\begin{figure*}[!t]
  \centering
  \includegraphics[width=\hsize]{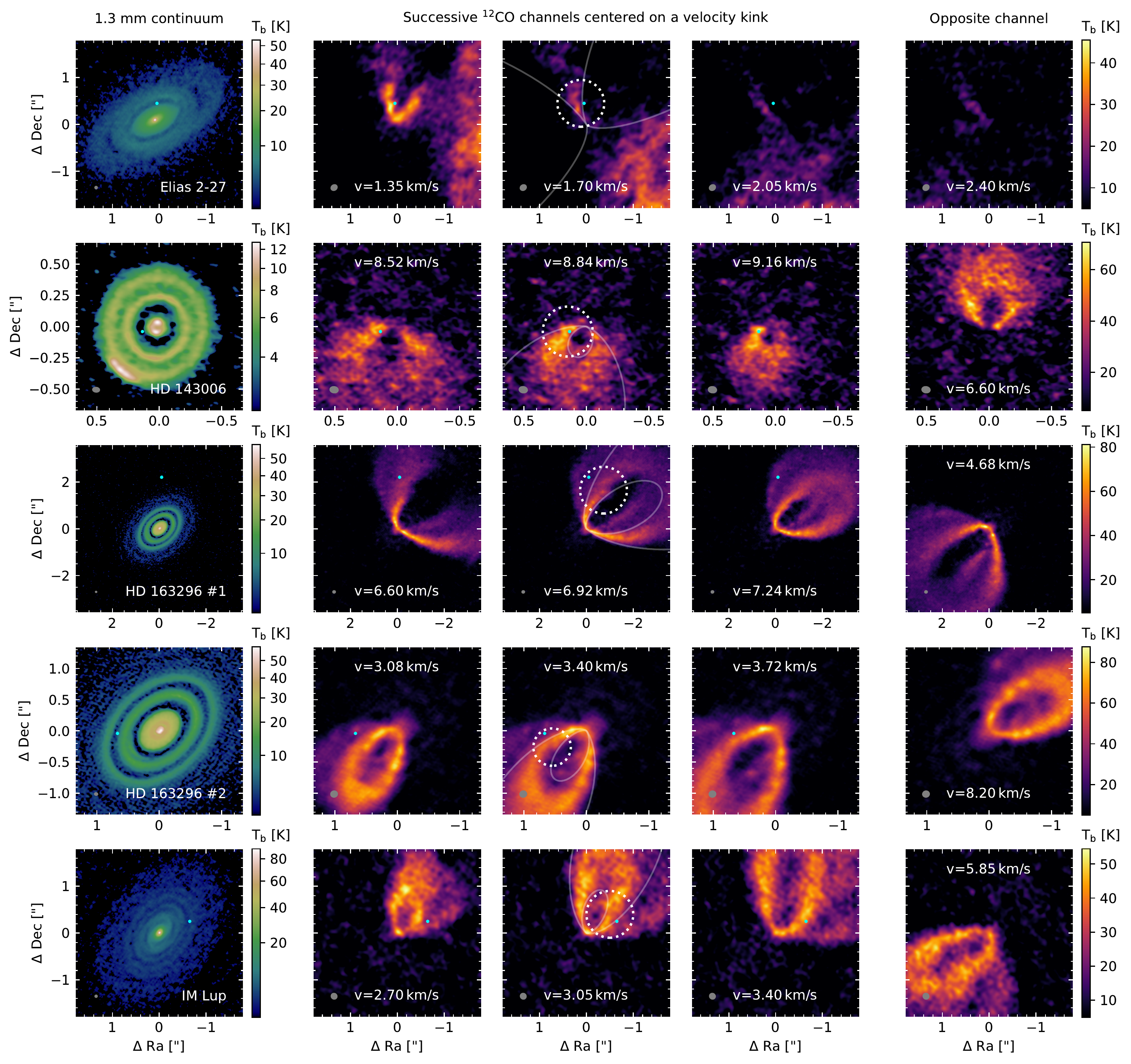}
  \caption{Candidate velocity kinks detected in the $^{12}$CO J=2-1 DSHARP data. Dashed circles indicate the velocity kinks, and the cyan dots the location of the planet assuming it is in the disk midplane. Solid lines in the third column indicate the expected location of the isovelocity curves at $\pm 0.2\,v_\mathrm{Kep} \sin i$, where $v_\mathrm{Kep}$ is the Keplerian velocity at the location of the planet. In all 8 disks the candidate planet lies within a continuum dust gap. Note that channel spacing is half of the spectral resolution due to Hanning smoothing, and adjacent channels are not independent. Strong cloud contamination is present for Elias 2-27.
\label{fig:kinks}}
\end{figure*}

\begin{figure*}[!t]
  \centering
  \includegraphics[width=\hsize]{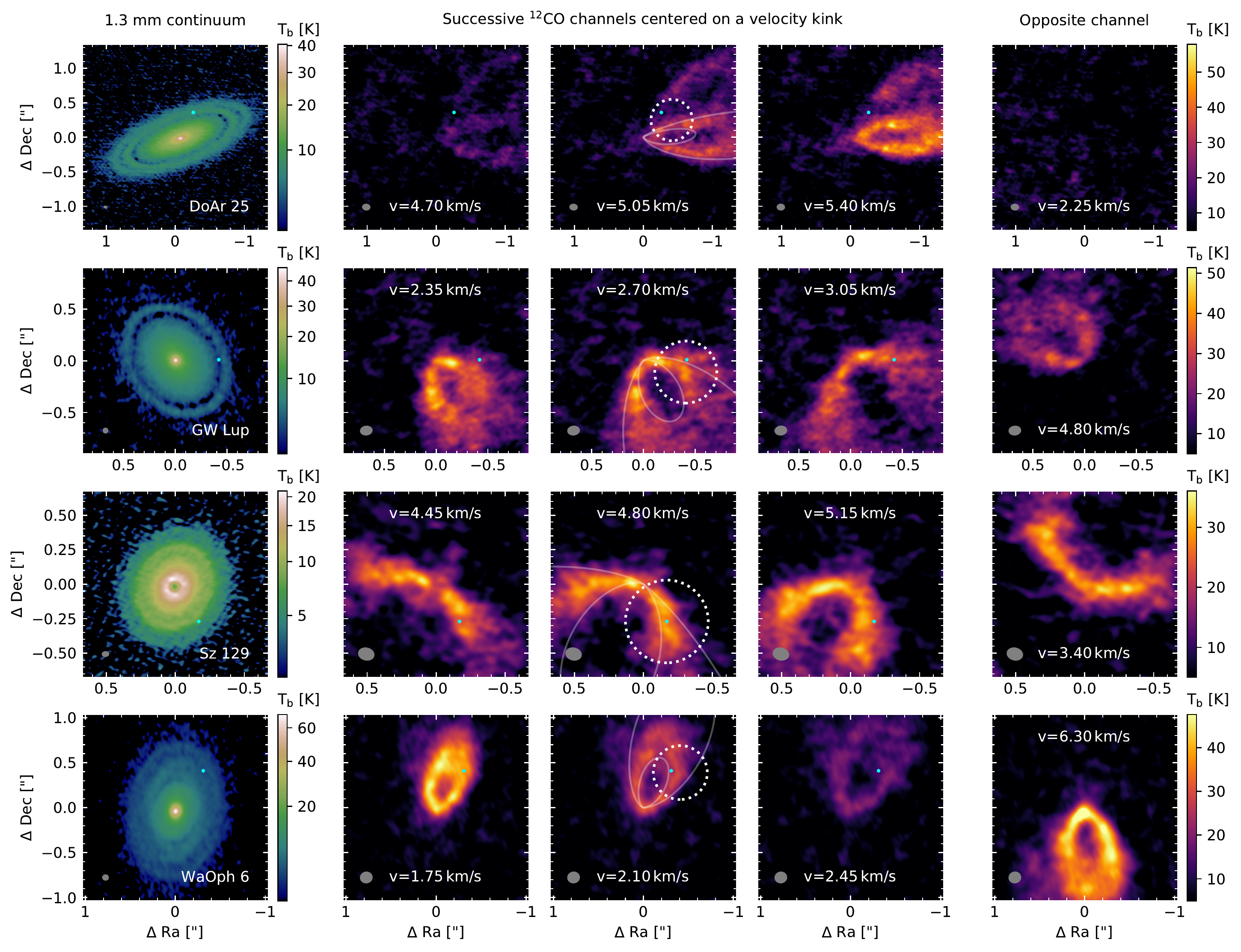}
  \caption{Additional potential velocity kinks.  The detections are not as clear as in Fig.~\ref{fig:kinks} due to the lower quality image reconstruction. Panels are the same as in Fig.~\ref{fig:kinks}. Strong cloud contamination is present for DoAR~25 and WaOph~6.\label{fig:kinks2}}
\end{figure*}

We detected the velocity kinks by manually inspecting the publicly available, science ready DSHARP $^{12}$CO J=2-1 data cubes\footnote{\url{https://almascience.org/alma-data/lp/DSHARP}}. We excluded HT~Lup and AS~205 from our analysis as they are binary stars, and display a perturbed velocity field.

Continuum subtraction can significantly affect the measured brightness temperature and apparent morphology of line emission when
the line is optically thick, and the line and continuum intensities are comparable \citep{Boehler17}.
To ensure that our kinks are not artificially created by continuum subtraction, we re-imaged all the CO data without continuum subtraction. For all disks, we used the released non-continuum-subtracted visibilities, and
adapted the released scripts to re-image with the same parameters as the fiducial (continuum subtracted) DSHARP images (see Table 5 in \citealp{Andrews2018b}).
The initial DSHARP data was imaged with a mix of manual masking and auto-masking. Here we used auto-masking in all cases, and for the disks that were imaged with manual masks in DSHARP, we also re-imaged the continuum subtracted data to ensure that there was no significant differences with the original CO cubes.
The kinks we present below are detected in both the continuum subtracted and non-continuum subtracted cubes in each case.

The channel width of the DSHARP data is 244\,kHz. Due to Hanning smoothing, the velocity resolution is 640\,m.s$^{-1}$.
We re-imaged the cubes with the same channel spacing as the publicly released calibrated visibilities and fiducial DSHARP CO cubes: 350\,m.s$^{-1}$, except for HD~163296 and HD~143006 which were imaged with a velocity spacing of 320\,m.s$^{-1}$ .

\section{Results}

We detect 9 velocity kinks in 8 of 18 selected DSHARP objects. Figures~\ref{fig:kinks} and ~\ref{fig:kinks2} show the continuum emission (left column) and three selected successive velocity channels (second, third and fourth columns, respectively) in the $^{12}$CO J=2-1 emission for our nine candidate CO emission kinks. The third column shows the channel where the velocity kink is most prominently detected, with the kink itself indicated by the dotted white circle. The last column shows the CO emission in the channel at opposite velocity, demonstrating that the kink is not an axisymmetric feature.
The kinks were identified by visual inspection of the CO cubes. There is currently no quantitative means of assessing confidence in a detection, when the signal-to-noise ratio is limited, and when background/foreground contamination exists as we are probing the disks in the $^{12}$CO line.

We have grouped the detections according to the visual quality of the detection:  \emph{firm} detections in Figure~\ref{fig:kinks} and \emph{potential} detections in Figure~\ref{fig:kinks2}. Independent observations will be required to confirm these candidate velocity kinks.

The DSHARP program was mostly aiming at high spatial resolution in continuum emission. The $^{12}$CO J=2-1 data only have a coarse spectral resolution of $\approx$\,640\,m.s$^{-1}$, and were imaged at 0.1'' spatial resolution due to the low signal-to-noise ratio \citep{Andrews2018b}. Most velocity kinks appear narrow in velocity, suggesting a local origin. The isovelocity curves overplotted in the central panel of Figures~\ref{fig:kinks} and \ref{fig:kinks2} show that in all cases, velocity perturbations are limited to 20\% of the local Keplerian velocity. Accross the kinks, the emission wiggles between the 2 isovelocity curves at $\pm 0.2\,v_\mathrm{Kep}$ without crossing them (except for HD~143006), indicating velocity deviations of order 10 to 15\,\%. The limited velocity resolution prevents unambiguous detections of the velocity kinks in every case. The kinks are seen to a lesser extent in the neighbouring channels (second and fourth columns in Fig.~\ref{fig:kinks} and \ref{fig:kinks2}).

Assuming the velocity kink is caused by a planet, the cyan dot in each panel indicates the inferred location of the protoplanet when the velocity kink is deprojected to the disk midplane (using the method described in \cite{Pinte2018a} to measure the CO emitting layer altitude). The location of the candidate protoplanets are indicated in Table~\ref{tab:protoplanets}. The detection in HD~143006 was already presented in \cite{Perez_DSHARP}, and the first one in HD~163296 in \cite{Pinte2018b} and \cite{Isella_DSHARP}, but we also include them here for completeness.

In every case the candidate protoplanet would lie within a gap in the dust continuum emission. In two cases (IM~Lup and WaOph~6) the protoplanet would also lie at the tip of a spiral arm detected in the dust continuum emission \citep{Huang2018b}. There is however no obvious correlation between the detection of a kink and spectral type, molecular cloud membership (as a proxy for age), or properties of the dust continuum gap.

\begin{table*}
\begin{center}    \begin{tabular}{|p{1.55cm}|p{1.3cm}|p{1.3cm}|p{0.8cm}|p{1.4cm}|p{1.4cm}|p{1cm}|p{1cm}|p{1cm}|p{1.3cm}|p{2cm}|}
\hline
Object & planet sep. ["] & planet PA [deg]& S/N$^a$ & Velocity width $\Delta v$ [m.s$^{-1}$]&  $\Delta v$ / $v_\mathrm{Kep}$ & Gap radius [au] & Gap width $\Delta/r$ & M$_{\rm star}$ [M$_\odot$] & Distance [pc] & Planet location \\
\hline
Elias 2-27 &   0.32 &      -6 &     12 & ? & ? &   69 & 0.18 & 0.49 & 116$^{+19}_{-10}$ & semi-minor axis\\
HD 143006 &   0.14 &   -107 &     10 & $\approx 700$ & $\approx 0.20$&  22 & 0.62 & 1.78 & 165 $\pm$ 5 & red-shifted near side\\
HD 163296 &   2.20 &     3 &     27 & $< 700$ & $< 0.26$ &  260 & - & 2.04 & 101 $\pm$ 2 & red-shifted near side\\
HD 163296 &   0.67 &   -93 &     36 & $\approx 700$ & $\approx 0.15$ &   86 & 0.17 & 2.04 & 101 $\pm$ 2 & blue-shifted near side\\
IM Lup &   0.70 &    69 &     14 &  $<700$ & $<0.24$&  117 & 0.13 & 1.12 & 158 $\pm$ 3 & blue-shifted near side\\
DoAr 25 &   0.44 &    36 &      7 & ? & ? &   98 & 0.15 & 0.95 & 138 $\pm$ 3 & red-shifted near side\\
GW Lup &   0.42 &    89 &     12 & $<700$ &  $ < 0.3$ & 74 & 0.15 & 0.46 & 155 $\pm$ 3 & blue-shifted near side\\
Sz 129 &   0.29 &   148 &     11 &  $<700$ & $<0.2$ &  64 & - & 0.83 & 161 $\pm$ 3 & red-shifted far side\\
WaOph 6 &   0.51 &    37 &     13 & ? & ? &   79 & - & 0.68 & 123 $\pm$ 2 & blue-shifted far side\\
\hline
\end{tabular}
\end{center}
$^{a}$ Signal-to-noise of the CO emission at the location of the velocity kink.
\caption{Summary of candidate protoplanets. Distances and stellar masses are as listed in \citet{Andrews2018b}. Gap widths are from \cite{Huang2018a}. The signal-to-noise is the ratio of the signal at the location of the kink divided by the RMS of the image far from the disk.
\label{tab:protoplanets}}
\end{table*}

Non-detections in the remaining 10 objects may be due to the limited signal-to-noise ratio and contamination of the $^{12}$CO emission by surrounding clouds. In particular, emission in Sz~114, SR~4, Elias~2-20, Elias~2-24, and WSB~52 is contaminated, and the signal-to-noise ratio is low for MY~Lup, HD~142666, DoAr~33.

AS~209 and RU~Lup are the only two sources that do not display any obvious CO emission kink despite modest cloud contamination and reasonable signal-to-noise ratio. The CO emission in RU Lup is contaminated by the outflow at large scales however, preventing any kink detection in the outer disk.

Perhaps surprisingly, we did not detect any obvious kink in AS~209, which is the DSHARP disk with the sharpest dust gaps (a small deviation might be present around $v=3.7$\,km.s$^{-1}$), as well as with a gas gap detected outside of the continuum disk \citep{Guzman2018}.

\section{Discussion}

\subsection{Origin of the velocity kinks}

Several observational effects and physical mechanisms may produce features in the channel maps that look like velocity kinks.

The most obvious one is the reconstruction process at low signal-to-noise ratio which often results in patchy emission that could be mistaken for kinks. We cannot exclude that such artefacts are present in the DSHARP data, but we indicate in Table~\ref{tab:protoplanets} the measured signal-to-noise ratio at the location of the kink. We also imaged the cubes with deeper CLEANing, as well as different choices of {\it uv} taper, robust parameters, CLEANing scales and velocity binning.  The various changes to the imaging parameters do not significantly affect the kinks (see Appendix~\ref{sec:imaging}). Deeper observations and in optically thinner tracers (less affected by cloud contamination) are needed to fully confirm these candidate detections.

Optical depth effects and in particular continuum subtraction may also affect the emission in a given channel, potentially mimicking a kink. Comparison of the continuum subtracted and non-continuum subtracted maps shows that this does not significantly affect our results.  All kinks are recovered in both sets of maps, at the same location.

A locally reduced altitude of the CO emitting layer, for instance due to an axisymmetric physical gap in the gas density structure, would also result in a displacement of the emission, but this distorted emission should be seen in all channels and should be associated with variations in brightness temperature.
This is for instance seen for HD~163296, where gaps have been previously detected in the gas \citep{Isella2016}, in particular South of the star in the channels presented in Fig.~\ref{fig:kinks}. The velocity kink we detect East of the central object displays a significantly different fork shape, but we cannot rule out that it is at least partly due to a CO line optical depth effect, rather than a velocity signature. Higher spectral resolution observation at a similar spatial resolution are necessary to conclusively distinguish between these two possibilities. Similarly, CO gaps have been detected by \cite{Favre2018} in the disk of AS~209 and may hide the presence of small velocity kinks. Additionally, \cite{Teague2018b} measured azimuthally averaged rotation curves in AS~209 and detected deviations from Keplerian rotation at the 5\% level. It remains unclear whether these deviations in the velocity profile reveal intrinsically azimuthally symmetric deviations or averaged localised deviations. As far as we can tell, there is no significant localised deviation in the velocity field of the disk as probed by $^{12}$CO.

\vspace{5mm}
Non-Keplerian motion may be unrelated to planet wakes. The deviation from Keplerian velocity may occur in the radial, vertical or azimuthal directions, or any combination thereof.

Spiral arms caused by (internal or external) companions more massive than a planet or gravitational instabilities will also generate velocity perturbations. For instance deviations from Keplerian velocities in HD~142527 \citep{Casassus2015,Price2018}, and more recently HD~100546 \citep{Perez2019}, likely reflect the presence of a massive  companion ($\approx$ 0.3\,M$_\odot$ for HD~142527 and $> 10\,$M$_\mathrm{Jup}$ for HD~100546). The main difference is that large-scale spirals will produce velocity deviations over a significant fraction of the disk and a significant range of velocities (for instance $\approx$ 7\,km.s$^{-1}$ for HD~100546, \citealp{Perez2019}), whereas embedded planets would only produce a localised velocity kink.

Pressure gradients at the edge of the gap could also cause non-Keplerian motions, however such perturbations occur over a wide range in azimuth, which is not seen for any of our candidate detections above.

Foreground extinction in $^{12}$CO sometimes makes it difficult to assess if velocity kinks seen in the DSHARP data are localised (in both space and velocity). We indicate in Table~\ref{tab:scoring} when this is the case. None of our detections are extended over a significant fraction of the disk, but some of them are strongly extincted or confused with cloud emission, preventing us from ruling out a large scale velocity feature. The most obvious cases are DoAr~25, Elias~2-27, and WaOph~6 where the channels adjacent to the main channel where the kink is detected are heavily extincted. If the cloud contamination is not uniform across the disk, this may artificially create an apparent kink by distorting the disk emission. We also detected hints of velocity deviations on the South side of Elias~2-27, but the $^{12}$CO emission is diffuse making it impossible to reach a definitive conclusion. More optically thin molecular lines should provide a more definitive answer.

Because the velocity deviations created by an embedded planet are small (around 10\% of the Keplerian velocity for a Jupiter mass planet at a few tens of au), a planet kink can often only be detected in a single channel at the DSHARP spectral resolution ($\approx$ 640\,m.s$^{-1}$), making it difficult to assess whether the detection is robust or whether it is affected by imaging artefacts. For HD~163296 and HD~97048 \citep{Pinte2018b,Pinte2019}, the high velocity resolution ($\approx$ 100 to 200\,m.s$^{-1}$) enabled detection of the kinks in several subsequent channels, providing for robust detections.

\subsection{A planetary origin?}

Based on the arguments presented above, for a velocity kink to be caused by a planet the perturbation should be, at minimum, i) detected in continuum and non-continuum subtracted data ii) detected at high signal-to-noise ratio iii) localised in velocity (e.g., to within 20\,\% of the local Keplerian velocity) and space (e.g., to within 3 beams) and iv) resolved in velocity (i.e. detected in at least 3 independent channels). In Table~\ref{tab:scoring} we assess our nine candidate kinks against these four criteria. Only the main kink seen in HD~163296 at 2.2'' satisfies all four criteria with the currently available datasets. The remaining uncertainty is due to the poor quality of the data rather than specifically ruling out a planetary origin.

\begin{table}
\begin{center}
\begin{tabular}{|p{1.9cm}|p{1.3cm}|p{1.cm}|p{1.3cm}|p{1.2cm}|}
  \hline
  Candidate kink & Indep. of cont.~sub.$^{a}$ &  S/N $>~10$ & Localised$^{b}$ & Resolved in $v^{c}$\\
  \hline
  Elias 2-27 &   \checkmark &  \checkmark &  {\bf ?} & \checkmark  \\
  HD 143006 &   \checkmark &  \xmark &   \checkmark & \checkmark  \\
  HD 163296 &   \checkmark &  \checkmark &   \checkmark & \checkmark \\
  HD~163296~\#2 &   \checkmark &  \checkmark &   \checkmark & \xmark  \\
  IM Lup &    \checkmark &  \checkmark &   \checkmark & \xmark  \\
  DoAr 25 &  \checkmark  & \xmark  &  {\bf ?} & {\bf ?}  \\
  GW Lup &    \checkmark &  \checkmark &   \checkmark & \xmark  \\
  Sz 129 &   \checkmark &  \checkmark &  \checkmark & \xmark  \\
  WaOph 6 &   \checkmark &  \checkmark &  {\bf ?} & {\bf ?}  \\
  \hline
\end{tabular}
\end{center}
$^{a}$ Detected in continuum subtracted \& non-subtracted data.\\
$^{b}$ Localised in space ($\lesssim$\,3\,beams) \& velocity ($\lesssim$\,20\% Keplerian velocity).\\
$^{c}$ Detected in at least 3 independent channels.\hfill
    \caption{Summary of candidate kinks assessed against our four criteria to assess a planetary origin.}
    \label{tab:scoring}
\end{table}

The most compelling argument towards a planetary origin is that, when deprojected, all nine kinks point to a perturber located in a continuum dust gap and, in two cases, at the tip of a spiral arm. There is an increasing consensus (mainly from theoretical modelling efforts) that dust gaps seen with ALMA are caused by embedded bodies \citep[e.g.][]{Dong2015,Bae2017,Zhang_DSHARP,Lodato2019}, so detection of perturbing bodies in ALMA kinematics is not unexpected. We emphasize that this deprojection was performed blind in the continuum subtracted channel maps, i.e. without reference to the dust gap locations \citep{Pinte2018a}.

\begin{figure*}[!t]
  \includegraphics[height=0.33\hsize]{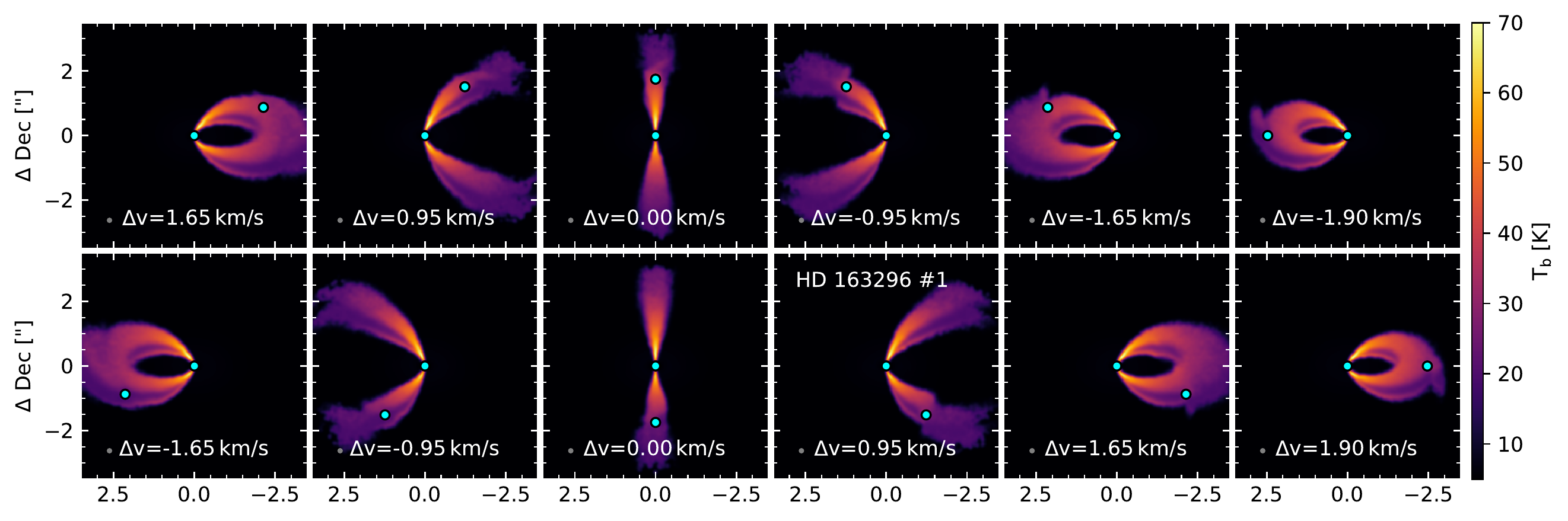}
  \includegraphics[height=0.343\hsize]{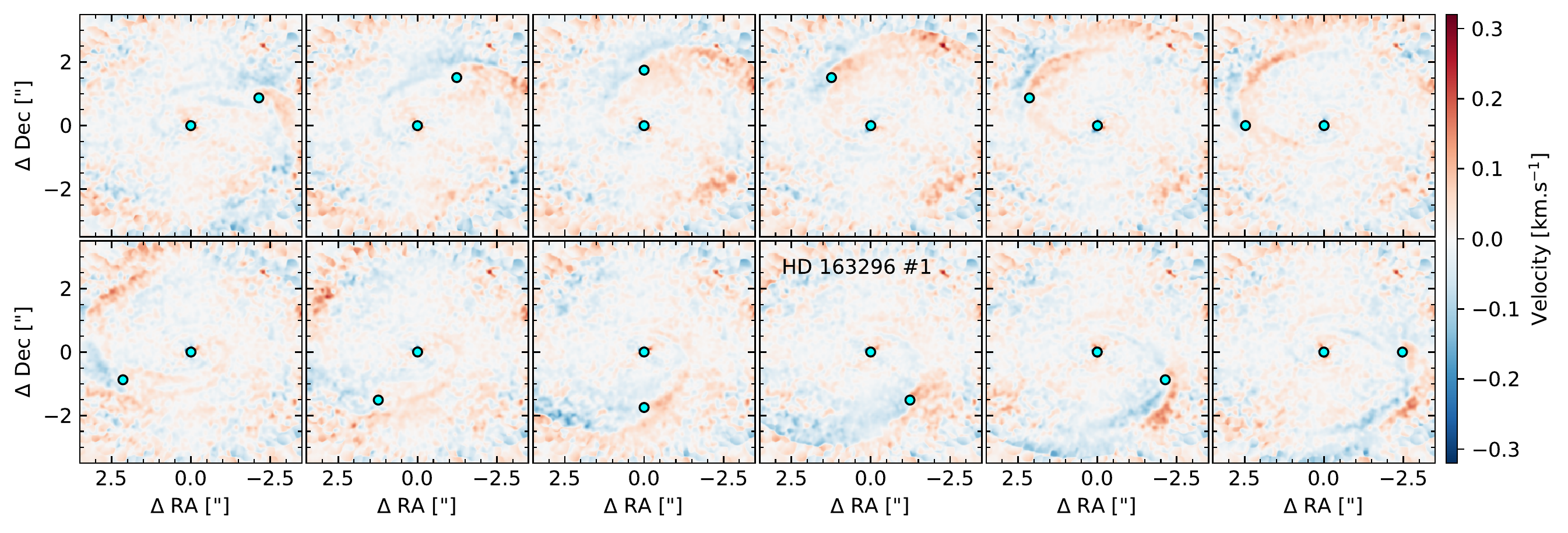}
  \caption{Detectability of a 3\,M$_\mathrm{Jup}$ planet-induced deviation from Keplerian rotation as a function of planet azimuth. Disk semi-major axis is horizontal with near side towards the bottom of each panel. Planet is rotated by 30$^\mathrm{o}$ between each panel. Models have been convolved by a Gaussian beam with full width at half maximum of 0.1". Sink particles are marked by a cyan dot. The labelled panel indicates the configuration of the main kink of HD~163296 (with a different PA). \emph{Top two rows:} $^{12}$CO synthetic channel maps with a width of 50\,m/s. In each panel, the velocity channel is selected to be the closest to the radial velocity of the planet. \emph{Bottom two rows:} differential rotation maps for the same planet positions. In each panel, we compute the first moment of the model from which we subtract an azimuthally averaged model. The amplitude of the colormap corresponds to the DSHARP data velocity resolution (640\,m/s). \label{fig:models}}
\end{figure*}

The velocity kinks we detected are not uniformly distributed in azimuth in the discs. In particular, we did not find any along the disk semi-major axis and only one along the semi-minor axis (Table~\ref{tab:protoplanets}). Foreground contamination,  when present, usually encompasses the systemic velocity, and often hides emission along the semi-minor axis. Additionally, the distortions in the channel maps depend on the inclination, distance and azimuth of the planets. Figure~\ref{fig:models} shows the detectability of a planet-induced velocity deviation as a function of the planet azimuth. We used the disk model presented in \cite{Pinte2018b} for HD~163296, with a 3\,M$_\mathrm{Jup}$ planet at 260\,au.  They appear stronger in the red-shifted half of the near side of the disk (as well as the blue shifted half of the far side), but appear with smaller amplitude on the other half of the disk. Velocity perturbations are most difficult to detect along the semi-major axis, where they appear as a small tail originating from the planet tracing the spiral arm. In this region of the disk, \emph{i.e.}, near the tip of the iso-velocity loop, emission is diffuse making it difficult to detect such a signal in actual data.

  \cite{Casassus2019} suggested that the location of the planet could be pinpointed by searching for ``Doppler flips'' in the rotation map of the disk, after subtraction of the Keplerian rotation. For each position of the planet in Fig.~\ref{fig:models}, we computed the moment 1 map and subtracted an azimuthally averaged model (implemented by rotating each particle in the smoothed particle hydrodynamics simulation by a random angle in the disk plane prior to performing the radiative transfer). The corresponding differential rotation maps are presented in the last two rows of Fig.~\ref{fig:models}. In each panel, there is a sign reversal as suggested by \cite{Casassus2019}, but the amplitude is small, with maximal deviations of about 150\,m/s. The  Doppler flip is also dependent on the planet azimuth, with blue and red-shifted sides varying in amplitude and size. Some additional velocity sign reversals are also sometimes seen  in other regions of the disk, potentially complicating the extraction of the planet position.
  The rotation maps presented in Fig.~\ref{fig:models} were generated with a perfect knowledge of the velocity field and taking into account the radiative transfer effects at the emitting surface.
  For actual data, the velocity field and altitude of the emitting layers have to be estimated. Uncertainties in the Keplerian velocity model that is subtracted may limit the detectability of Doppler flips to high signal-to-noise cases.
  In particular, the predicted Doppler flip for the orientation of HD~163296 (labeled panel in Fig.~\ref{fig:models}) shows an amplitude twice smaller ($\approx 150\,m/s$) than the residuals in the differential rotation map presented in \cite{Casassus2019} ($\approx 300m/s$). This explains why they did not detect the counterparts of the kink as a Doppler flip in the first moment map. The absence of visible Doppler flip corresponding to the kink in HD~97048 is also not surprising given the residuals in the differential rotation maps presented by \cite{Casassus2019}.

Because of the tentative nature of the detections and the need for higher resolution observations, we are not yet confident enough to estimate planet masses from the candidate kinks seen in the DSHARP data. Detailed hydrodynamical and radiative transfer modelling of the kink candidates is currently the only way to determine planet masses from velocity deviations in a quantitative manner but would result in poorly constrained estimates given the low SNR and poor velocity resolution of the data.
We estimated the velocity deviations from the number of channels in which the kinks are detected.
The limited spectral resolution prevents us from measuring velocity deviations smaller than 640\,m.s$^{-1}$. By comparing with the expected location of the isovelocity curves if the disks were in Keplerian rotation, we estimate the observed perturbations to be of order  10-15\% of the local Keplerian velocity (Table~\ref{tab:protoplanets}).
From our previous modelling of velocity kinks, we expect masses of order 1--3 $M_{\mathrm{Jup}}$. In HD~143006, the velocity deviation appears significantly larger, pointing towards a more massive planet.

With the exception of HD~143006, such masses are larger by a factor 4 to 10 than the masses derived from the continuum gaps by \cite{Zhang_DSHARP} or \cite{Lodato2019}.
We found a similar discrepancy for HD~97048 \citep{Pinte2019}, where to get a coherent match to both the continuum and line data, our models required the dust grains dominating the emission to have a Stokes number of a few $10^{-2}$, suggesting very porous and/or fluffy aggregates.

The high detection rate of velocity kinks at several tens of astronomical units in the DSHARP dataset may appear in contradiction with the occurrence rate of known giant extrasolar planets at large distance from their host stars: in the range of a few up to 5\% for massive planets (M $>$ 2\,M$_\mathrm{Jup}$) beyond 10-20\,au \citep{Vigan2017,Nielsen2019}. This occurrence rate suffers from large uncertainties however, depending on the assumption of a hot- or cold-start model for the planet. For instance, using cold-start models, \cite{Stone2018} find that up to 90\% of FGK systems can host a 7-10\,M$_\mathrm{Jup}$ planet from 5 to 50\,au.
Part of this discrepancy could also reflect the selection biases of the DSHARP program, which was aimed at millimeter bright protoplanetary disks. This could in turn be biased towards objects that have formed massive planets at large radii.
Additionally, the DSHARP planets will likely migrate inward by the time the disk dissipates. Assuming one planet per gap and modelling the planet orbital evolution and accretion,  \cite{Lodato2019} found that the final distribution of the planets from the DSHARP sample (as well as the Taurus survey, \citealp{Long2018}) is consistent with the known properties of the exoplanet population, and would represent a good match to the distribution of cold Jupiters.

\section{Conclusions}
\begin{enumerate}
\item We found 9 localised (channel-specific) velocity perturbations indicative of non-Keplerian motion in DSHARP observations of 8 protoplanetary disks, out of the 18 selected sources.

\item When deprojected, we find that the velocity kinks are systematically associated with gaps seen in continuum emission, suggesting they share a common origin. The presence of embedded planets would naturally explain both the continuum rings and gas velocity deviations from Keplerian rotation.

\item If planets are indeed responsible for these tentative velocity kinks, they should have masses of the order of a Jupiter mass. This is 4 to 10 times higher than the estimates from the width and depth of the continuum gaps.

\item Limited spectral resolution and signal-to-noise ratio, as well as cloud contamination prevents us from reaching definitive conclusions in several cases. In particular, non-detections in other disks or in other gaps in disks where we detected a kink do not necessarily imply the absence of Jupiter mass planets.

\item {Synthetic models indicate that the shape and amplitude of the planet velocity kink in the channel maps and rotation maps depend on the system geometry, inclination, azimuth and distance of the planet. In particular, for a given planet, signatures appear fainter on the blue-shifted half of the near side, and red-shifted half of the far side of the disk, as well as along the disk semi-major axis. This suggests additional planet signatures could be found with higher signal-to-noise data.}

\end{enumerate}

High signal-to-noise follow-up mapping at similar spatial resolution to the DSHARP data ($\approx$ 0.1'') and at high spectral resolution ($\approx$ 100 m.s$^{-1}$) can be reached with ALMA for less abundant molecules, i.e. with less foreground/background contamination, than $^{12}$CO  with integration times ranging from 2-4h ($^{13}$CO) to 10-20h (e.g. C$^{18}$O, HCO$^+$).
Such observations would confirm the tentative detections presented above, and enable to characterise the embedded planets.

 \section*{Acknowledgments}

 This Letter makes use of the following ALMA data:
 and ADS/JAO.\-ALMA\#2016.1.00484.L. ALMA is a partnership of ESO (representing
 its member states), NSF (USA) and NINS (Japan), together with NRC (Canada),
 MOST and ASIAA (Taiwan), and KASI (Republic of Korea), in cooperation with the
 Republic of Chile. The Joint ALMA Observatory is operated by ESO, AUI/NRAO and
 NAOJ. The National Radio Astronomy Observatory is a
facility of the National Science Foundation operated under cooperative
agreement by Associated Universities, Inc.
C.P., D.J.P. and V. C. acknowledge funding from the Australian Research Council via
FT170100040, FT130100034, and DP180104235. C.P., F.M, G.vdP. and Y.B, acknowledge funding
from ANR of France (ANR-16-CE31-0013). S.M.A. acknowledges support from the National Aeronautics and Space Administration under grant No.~17-XRP17\_2-0012 issued through the Exoplanets Research Program. L.P. acknowledges support from CONICYT project Basal AFB-170002 and from FONDECYT Iniciaci\'on project \#11181068.

\vspace{5mm}
\facilities{ALMA.}

\software{{\sf CASA} \citep{McMullin07}, {\sf phantom} \citep{Price2018}, {\sf mcfost} \citep{Pinte06,Pinte09}.}

\bibliographystyle{aasjournal}
\bibliography{biblio}

\listofchanges

\appendix
\section{Imaging tests}
\label{sec:imaging}

\begin{figure*}[!h]
  \centering
  \includegraphics[width=\hsize]{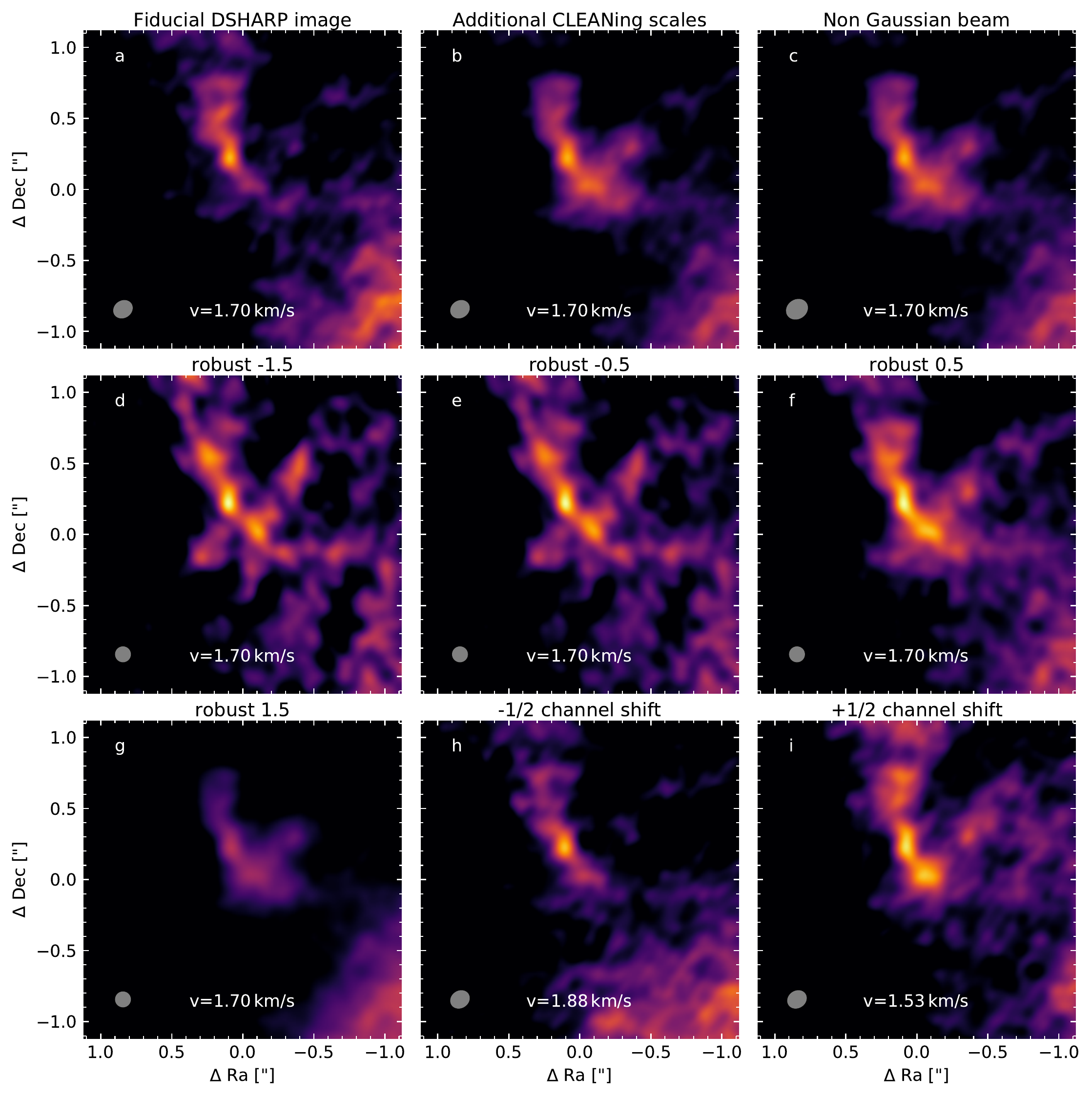}
  \caption{Imaging of Elias 2-27 with different parameters. a. Fiducial DSHARP image (but without continuum subtraction). b. Multi-scale CLEANing was performed with scales of 0, 5, 10, 20, 40, 80, 160, 320, instead of 0, 10, 25, 75, 150.   c. Image generated by convolving the CLEAN model with a Gaussian beam of same area as the dirty beam to take into account the wings of the beam generated by the combination of different ALMA configurations. d to g. CLEANing was performing with increasing values of the robust parameter. Images were then convolved by a Gaussian beam to reach the same 0.1'' spatial resolution.  h. and i. Imaging was performed with the same setup as the fiducial image but at velocities shifted by 1/2 of a channel ($\approx$ 1/4 of the velocity resolution).\label{fig:imaging_tests_Elias_2-27}}
\end{figure*}

\begin{figure*}[!h]
  \centering
  \includegraphics[width=\hsize]{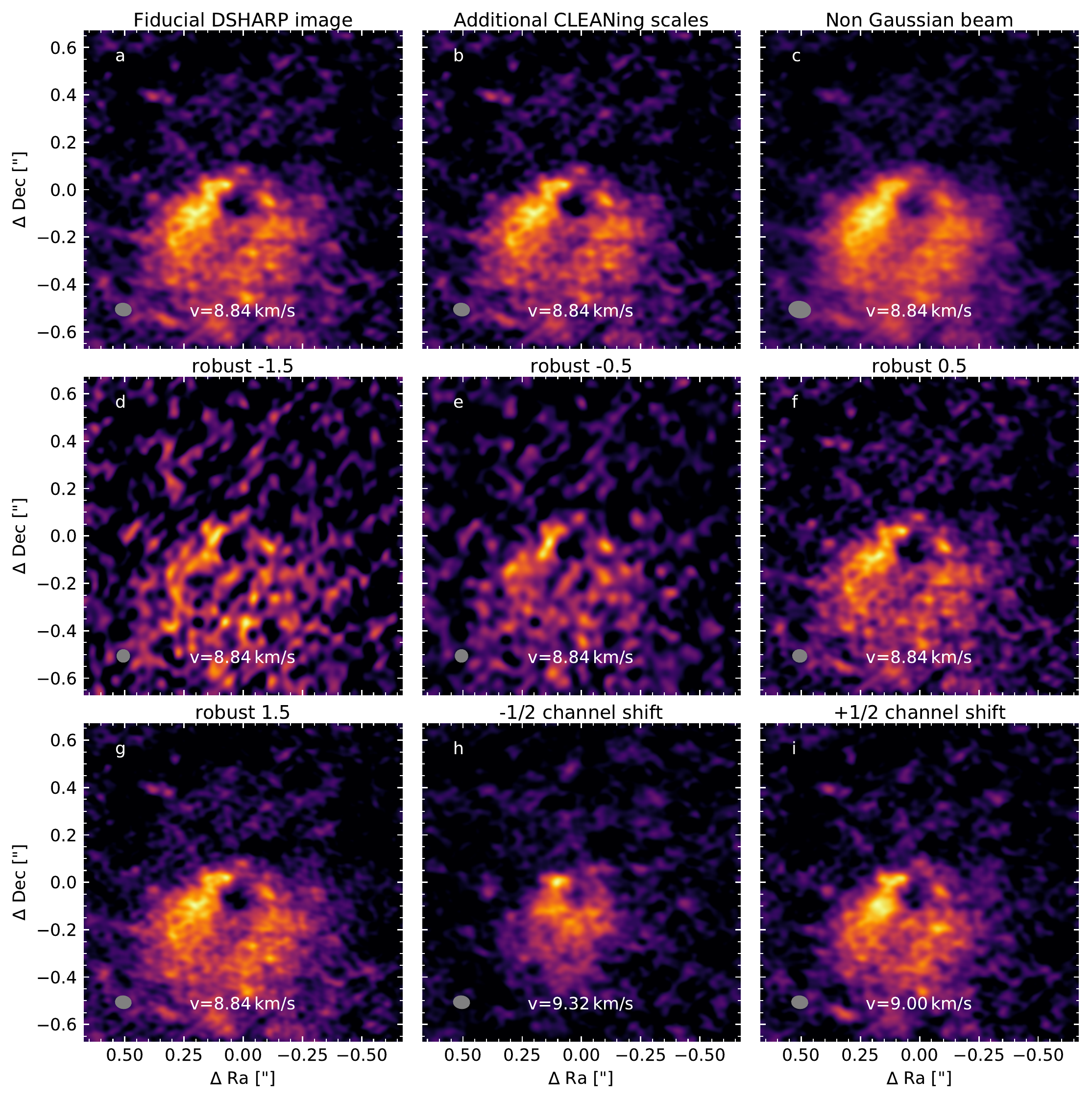}
  \caption{Same as Figure~\ref{fig:imaging_tests_Elias_2-27} but for HD~143006.\label{fig:imaging_tests_HD143006}}
\end{figure*}

\begin{figure*}[!h]
  \centering
  \includegraphics[width=\hsize]{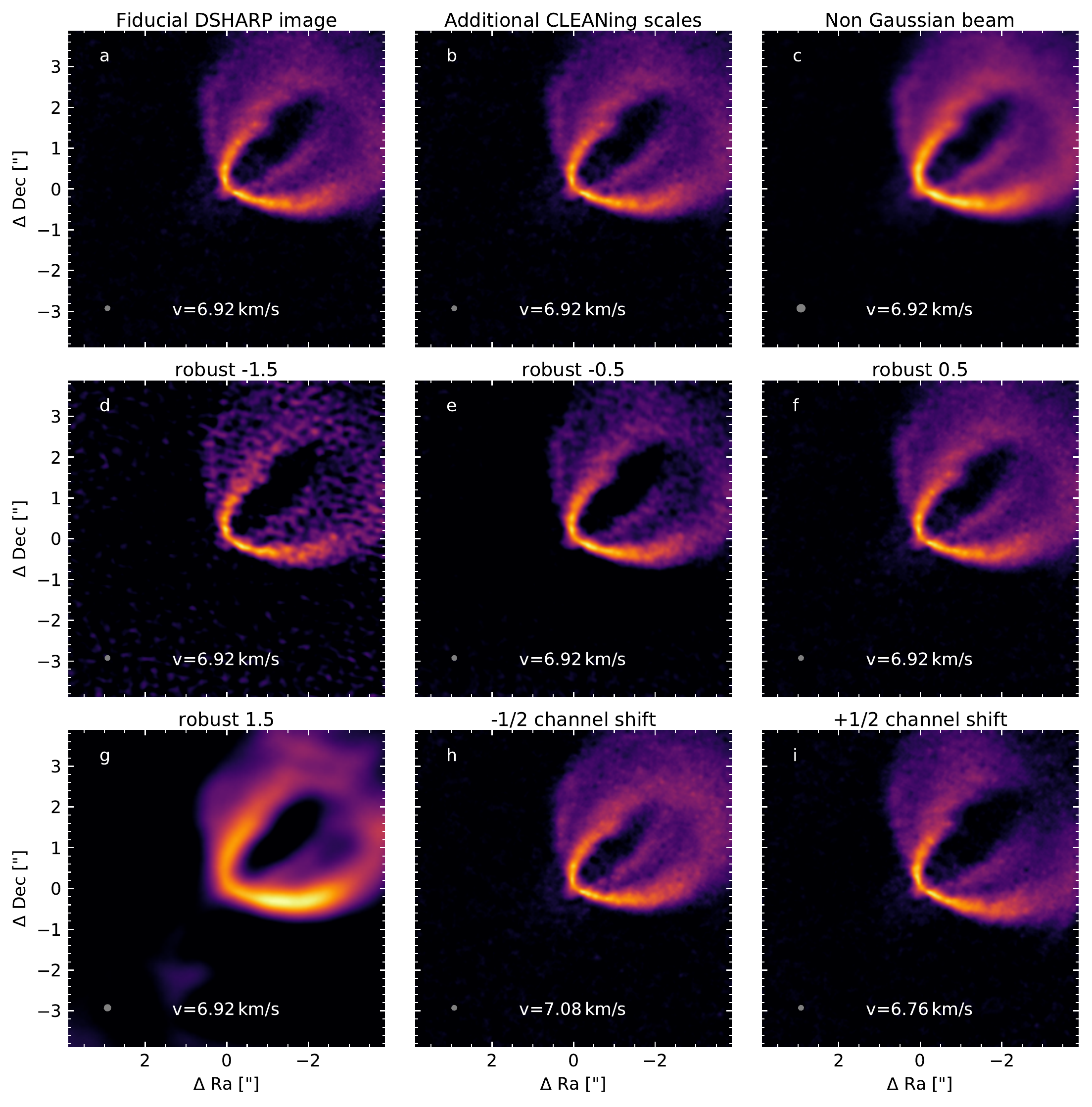}
  \caption{Same as Figure~\ref{fig:imaging_tests_Elias_2-27} but for HD~163296 \#1.\label{fig:imaging_tests_Elias_HD163296_1}}
\end{figure*}

\begin{figure*}[!h]
  \centering
  \includegraphics[width=\hsize]{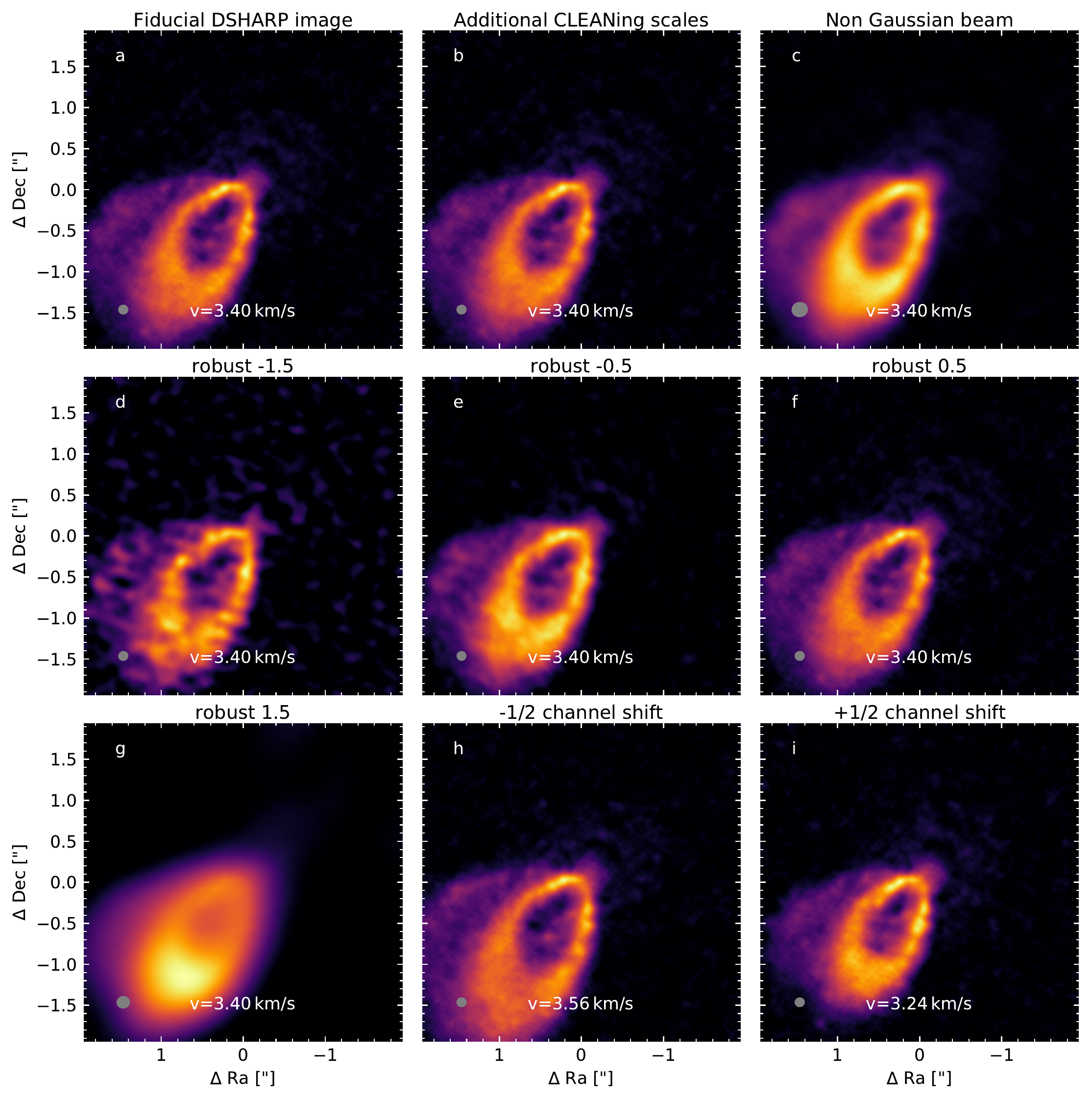}
  \caption{Same as Figure~\ref{fig:imaging_tests_Elias_2-27} but for HD~163296 \#2.\label{fig:imaging_tests_Elias_HD163296_2}}
\end{figure*}

\begin{figure*}[!h]
  \centering
  \includegraphics[width=\hsize]{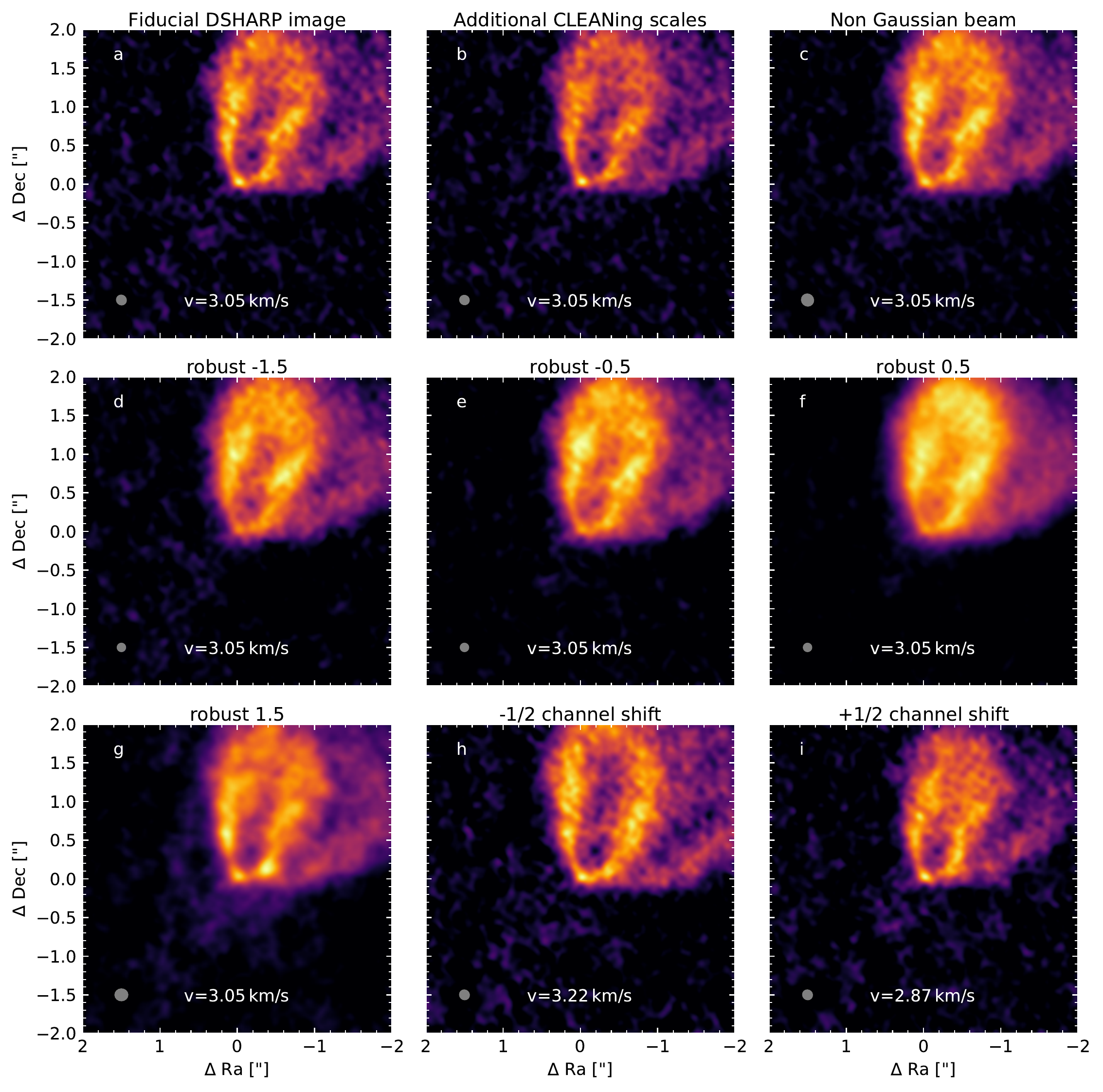}
  \caption{Same as Figure~\ref{fig:imaging_tests_Elias_2-27} but for IM~Lup.\label{fig:imaging_tests_IMLup}}
\end{figure*}

\begin{figure*}[!h]
  \centering
  \includegraphics[width=\hsize]{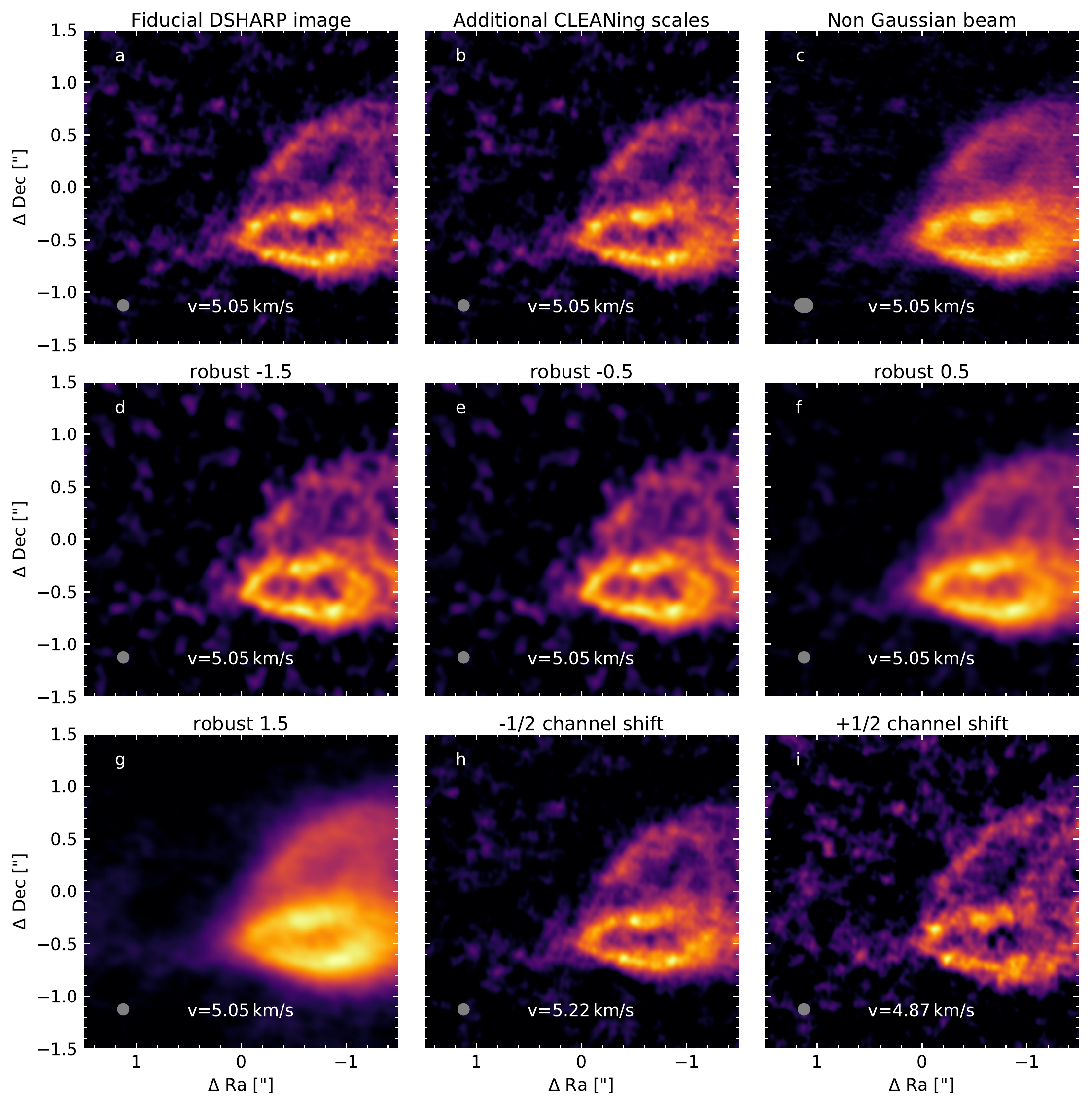}
  \caption{Same as Figure~\ref{fig:imaging_tests_Elias_2-27} but for DoAr~25.\label{fig:imaging_tests_DoAr25}}
\end{figure*}

\begin{figure*}[!h]
  \centering
  \includegraphics[width=\hsize]{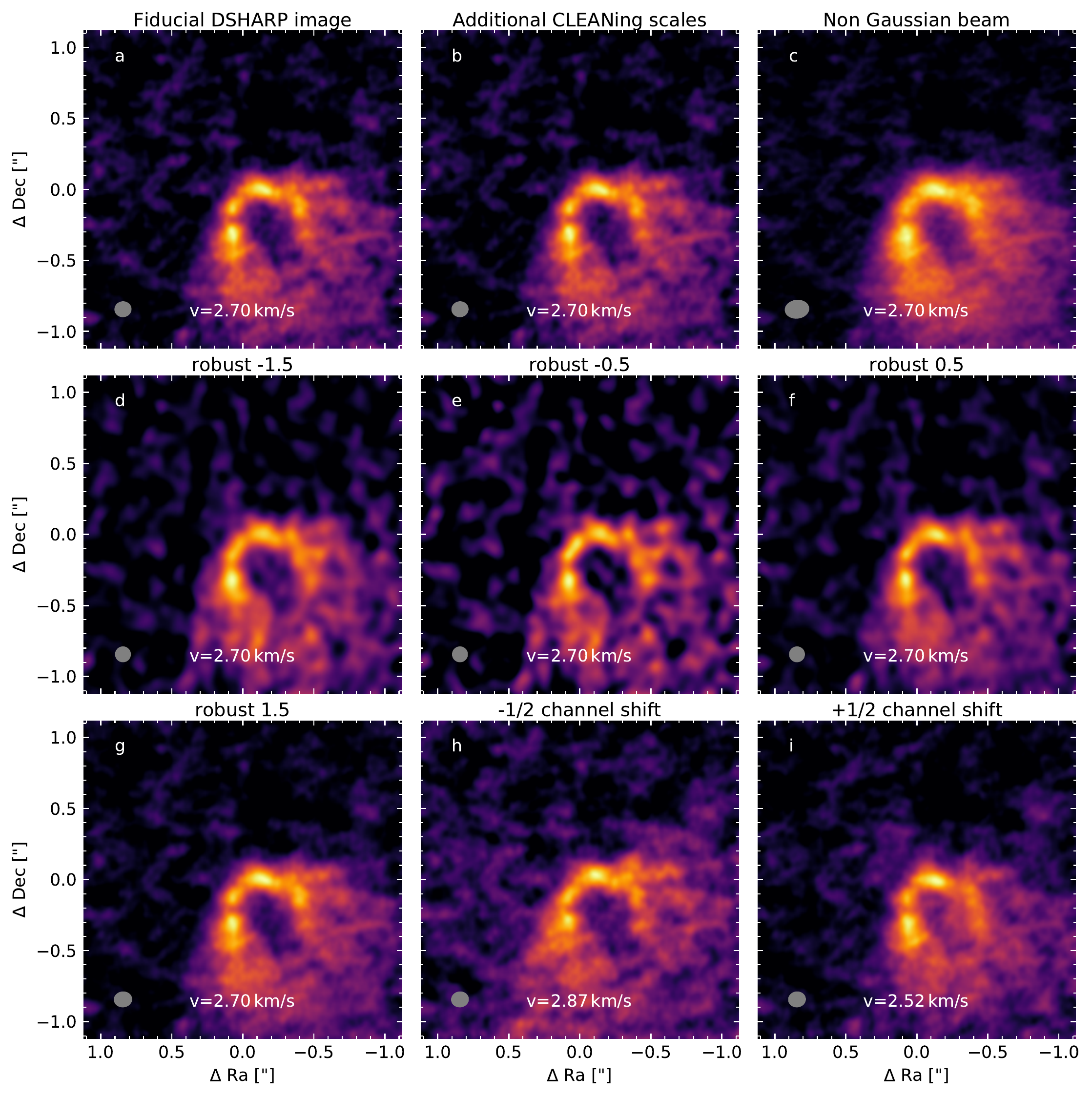}
  \caption{Same as Figure~\ref{fig:imaging_tests_Elias_2-27} but for GW~lup.\label{fig:imaging_tests_GWLup}}
\end{figure*}

\begin{figure*}[!h]
  \centering
  \includegraphics[width=\hsize]{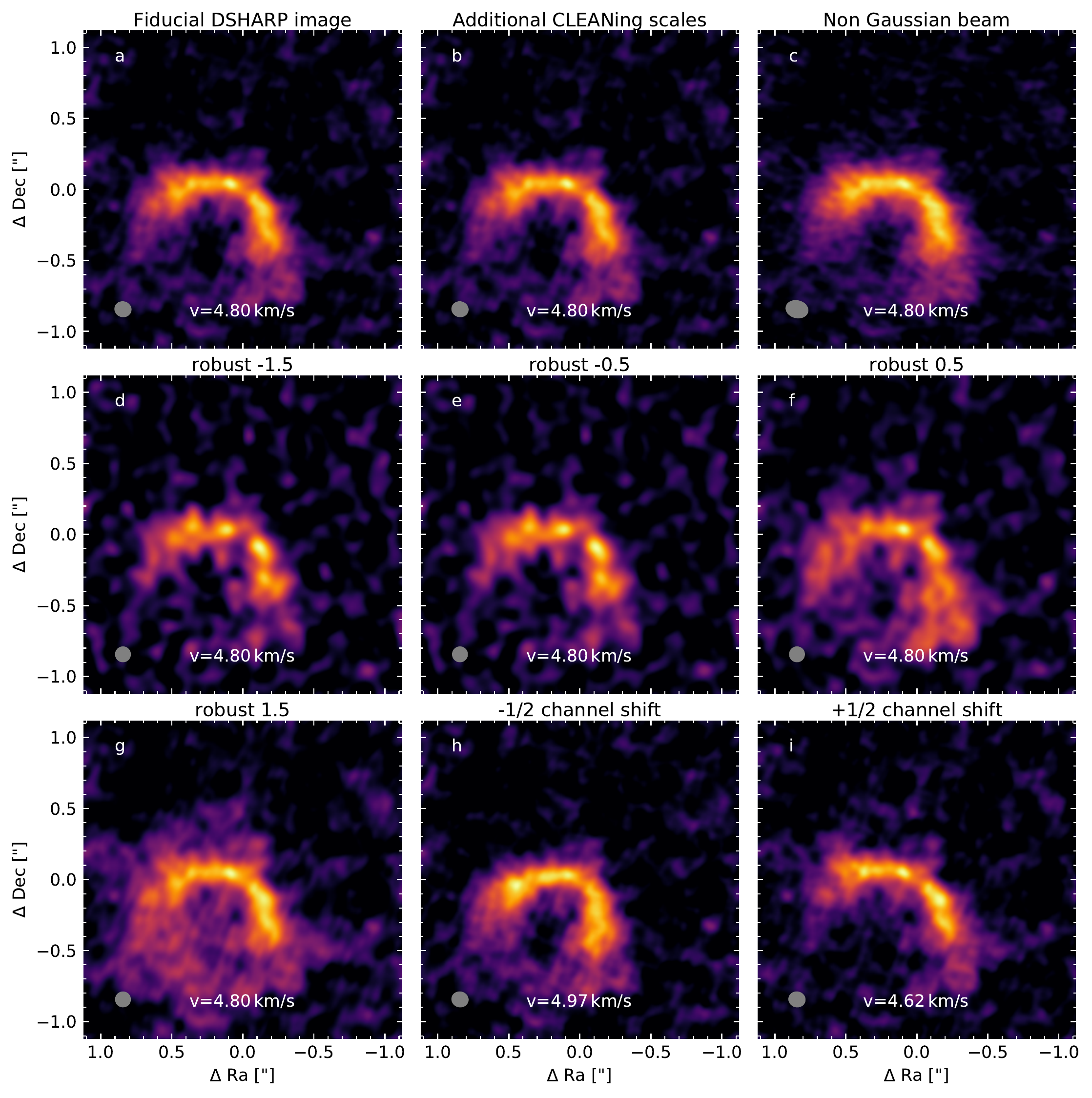}
  \caption{Same as Figure~\ref{fig:imaging_tests_Elias_2-27} but for Sz~129.\label{fig:imaging_tests_Sz129}}
\end{figure*}

\begin{figure*}[!h]
  \centering
  \includegraphics[width=\hsize]{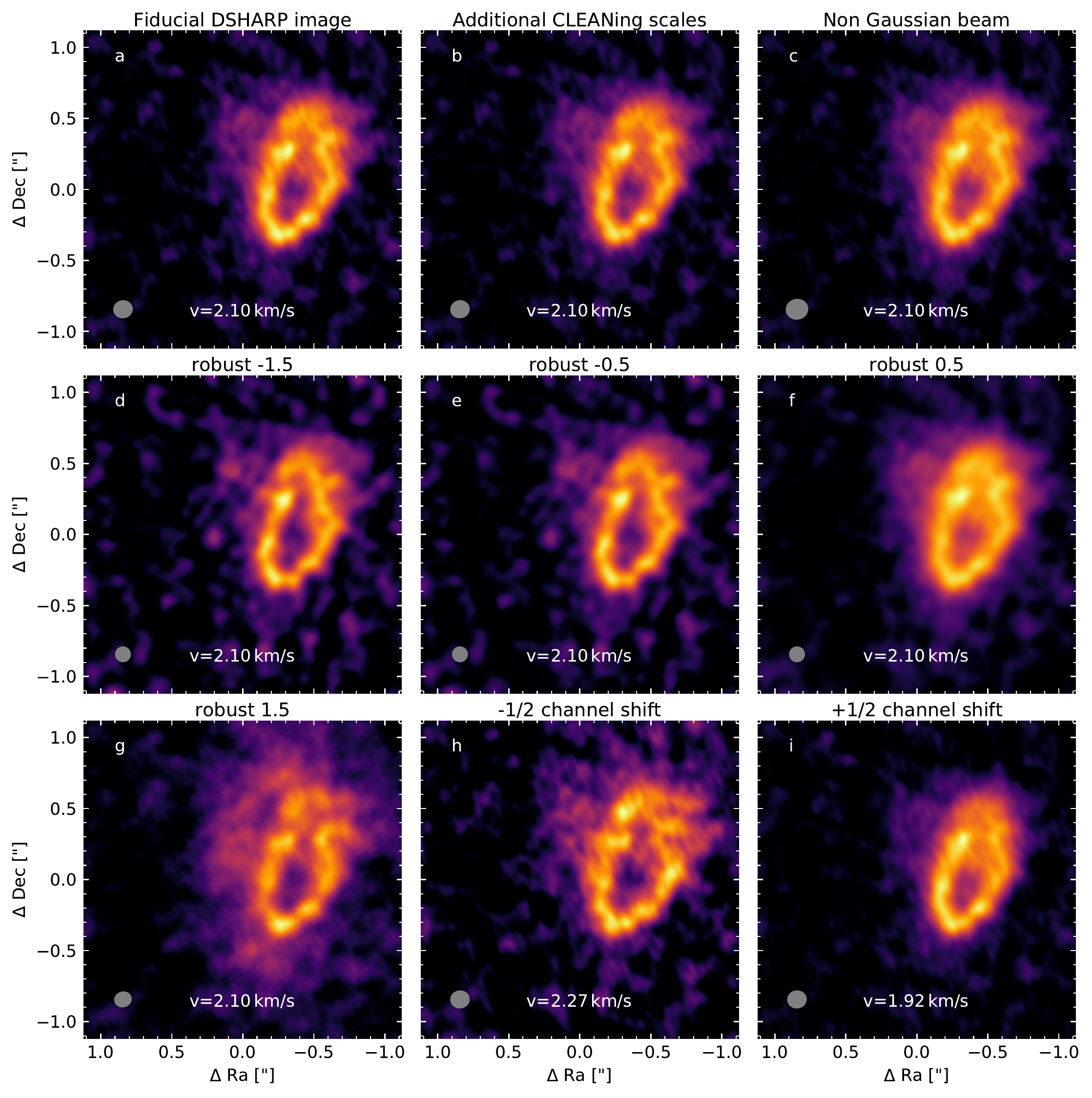}
  \caption{Same as Figure~\ref{fig:imaging_tests_Elias_2-27} but for WaOph~6.\label{fig:imaging_tests_WaOph6}}
\end{figure*}

Given the low signal-to-noise and limited $uv$ coverage of the DSHARP CO  observations (which were designed with the continuum as primary goal), the synthesis imaging limitations can potentially create artefacts that mimic a velocity kink. Figures~\ref{fig:imaging_tests_Elias_2-27} to \ref{fig:imaging_tests_WaOph6} illustrate the imaging tests we have performed. In all cases, the velocity kink remains detected.

The discontinuities in the gridded $uv$ weight density when imaging the combined data from several ALMA configurations result in non-Gaussian beams with significant ``shelves'' \citep[e.g.][]{Jorsater95}. This may affect the flux of the image as the clean model is convolved with the clean beam (units Jy/clean beam), while the dirty image and residuals are both in units of Jy/dirty beam. When the beam is Gaussian, these beam areas are equivalent, but when the beam has shelves, the integrated areas quickly diverge. In the case of the DSHARP data, the beam areas differ by about 20\,\%. Panel c of Figures~\ref{fig:imaging_tests_Elias_2-27} to \ref{fig:imaging_tests_WaOph6} shows an image where we convolved the clean model with a beam which matches the area of the dirty beam.

Narrow velocity deviations can be washed out if they are separated over 2 velocity bins.  We re-imaged the cubes with an offset of half a frequency bin (\emph{i.e.} 1/4 of the spectral resolution) compared to the fiducial DSHARP cubes (e.g. Fig.~\ref{fig:imaging_tests_Elias_2-27} to \ref{fig:imaging_tests_WaOph6}, panels h and i). In all cases, we detect the velocity kinks in the ``shifted'' cubes with a similar significance, except for IM~Lup where the kink appears sharper than in the fiducial DSHARP cube (Fig.~\ref{fig:imaging_tests_IMLup}).

\end{document}